\documentclass[aps,twocolumn,showpacs, superscriptaddress, groupedaddress,nofootinbib]{revtex4}  % for review and submission
\usepackage{graphicx}  % needed for figures
\usepackage{dcolumn}   % needed for some tables
\usepackage{bm,relsize}        % for math
\usepackage{amssymb, amsmath}
\usepackage{wasysym}
\usepackage{slashed}
\usepackage{lipsum, color}
\usepackage[usenames,dvipsnames,svgnames]{xcolor}
\usepackage{hyperref}
\hypersetup{colorlinks, citecolor=black, linkcolor=black, urlcolor=black}

\def\beq{\begin{equation}}
\def\eeq{\end{equation}}
\begin{document}
% The following information is for internal review, please remove them for submission
\widetext
\leftline{DESY 18-183}
\leftline{KEK-TH-2084}
%\centerline{NOT FOR PUBLIC DISTRIBUTION}

% the following line is for submission, including submission to the arXiv!!
%\hspace{5.2in} \mbox{Fermilab-Pub-04/xxx-E}

%\title{Signs of the R-axion}
\title{
New Axion Searches at Flavor Factories
%diphoton resonances below the Higgs mass at the LHC
}
\author{Xabier Cid Vidal}
\affiliation{Instituto Galego de F\'isica de Altas Enerx\'ias (IGFAE), Santiago de Compostela, Spain}
\author{Alberto Mariotti}
\affiliation{Theoretische Natuurkunde and IIHE/ELEM, Vrije Universiteit Brussel, and International Solvay
Institutes, Pleinlaan 2, B-1050 Brussels, Belgium}
\author{Diego Redigolo}
\affiliation{Raymond and Beverly Sackler School of Physics and Astronomy, Tel-Aviv University, Tel-Aviv
69978, Israel}
\affiliation{School of Natural Sciences, Institute for Advanced Study, Einstein Drive,
Princeton, NJ 08540, USA}
\affiliation{Department of Particle Physics and Astrophysics, Weizmann Institute of Science, Rehovot 7610001,Israel}
\author{Filippo Sala}
\affiliation{DESY, Notkestra{\ss}e 85, D-22607 Hamburg, Germany}
\author{Kohsaku Tobioka}
\affiliation{Department of Physics, Florida State University, Tallahassee, FL 32306, USA}
\affiliation{Theory Center, High Energy Accelerator Research Organization (KEK), Tsukuba 305-0801, Japan}
%\author{Xabier Albert and Sean?}
\date{\today}

%We use LHCb public diphoton data to derive a new bound on pp-produced resonances of masses between 4.9 and 6.3 GeV. This bound sets the current best limit on axion-like particles that couple to photons and gluons.
%Future LHCb dedicated searches would test masses from ∼XX to ∼YY GeV and ALP decay constants of a few TeV, a range motivated both from solutions to the strong CP-problem and from Dark Matter. Finally we derive the impact of ALP searches in Υ decays at BABAR and Belle-II and in B decays at Belle-II, showing that they would constitute the strongest probes for ALP masses lighter than ZZ GeV.

\begin{abstract}
\noindent We assess the impact of searches at flavor factories for new neutral resonances that couple to both photons and gluons. These are well motivated by ``heavy axion'' solutions of the strong CP problem and by frameworks addressing both Dark Matter and the Higgs hierarchy problem.  We use LHCb public diphoton data around the $B_s$ mass to derive the current best limit on these resonances for masses between 4.9 and 6.3 GeV.  We estimate that a future LHCb dedicated search would test an axion decay constant of $O$(TeV) for axion masses in the few-to-tens of GeV, being fully complementary to the low mass ATLAS and CMS searches. We also derive the impact of BABAR searches based on $\Upsilon$ decays and the future Belle-II reach.
\end{abstract}

\maketitle

\section{Introduction}

The lack of new physics at the Large Hadron Collider (LHC) and the lack of direct detection signal of dark matter (DM) at present experiments make it  necessary to rethink the theoretical questions in the SM from a wider viewpoint and trigger broader experimental searches for new physics (NP). In this paper we make a step in this direction by presenting a NP case for flavor factories at the intensity frontier. These are light resonances below the EW scale which are neutral under the SM gauge group and couple to both gluons and photons.
We show that flavor experiments have an unexploited potential to probe these states in a complementary mass range to previously proposed low-mass resonance searches at ATLAS and CMS~\cite{Mariotti:2017vtv}. Pointing out these gaps in the search program at flavor facilities is now a particularly important question in view of the upcoming LHCb upgrade and the Belle II data taking. 

The possibility we  consider here is that the new physics scale $M_\mathsmaller{\rm NP}$ lies beyond the reach of the LHC.
If that is the case, NP signals might still arise from pseudo-Nambu-Goldstone bosons (pNGBs) associated to spontaneously broken approximate symmetries.
These are often called axion-like particle (ALP) in the literature, they can be sensibly lighter than the NP scale ($m_a\ll M_\mathsmaller{\rm NP}$) and their couplings to the SM are controlled by the inverse of the decay constant $1/f$. Generically, one has $M_\mathsmaller{\rm NP}=g_{\ast} f$ with $g_{\ast}$ being the typical size of the couplings in the NP sector, so that probing weak enough couplings of the pNGB gives an indirect probe of the scale of new physics. 

The focus of this paper will be on pNGBs with $m_a$ between 2 and 20~GeV, a mass window within the reach of flavor experiments.  The driving question is 
\emph{whether flavor experiments can be sensitive to couplings of pNGBs small enough to probe new physics beyond the LHC reach.} This question has been partially addressed for ALPs which couple to the SM by mixing with the Higgs sector \cite{Wilczek:1977zn,Freytsis:2009ct} but it is surprisingly unexplored for ALPs with only gluon and photon couplings.  

In the large theory space of all the possible couplings of the ALP to the SM, having a non-zero coupling to gluons is particularly well motivated from the theory perspective. In this paper we will discuss in detail two particularly compelling examples: ``heavy'' QCD axions~\cite{Rubakov:1997vp,Berezhiani:2000gh,Hook:2014cda,Fukuda:2015ana,Dimopoulos:2016lvn,Holdom:1982ex,Choi:1988sy,Holdom:1985vx,Dine:1986bg,Flynn:1987rs,Choi:1998ep,Agrawal:2017ksf,Gaillard:2018xgk} and the $R$-axion~\cite{Nelson:1993nf,Goh:2008xz,Bellazzini:2017neg} in low energy SUSY-breaking.
%\AM{Do we want to modify the previous sentence and mention also the DM story? I think this sentence is old.}
As we will show, in these two classes of models the gluon coupling is unavoidable, the photon coupling generic, the mass range of interest for this paper can be easily achieved.
A TeV decay constant is theoretically favoured by ensuring the quality of the axion potential \cite{Kamionkowski:1992mf,Holman:1992us,Barr:1992qq,Ghigna:1992iv} or by explaining the DM relic abundance via thermal freeze out.
Besides these two examples, ALPs with both gluon and  photons couplings arise for instance as new pions in Composite Higgs models~\cite{Ferretti:2013kya,Belyaev:2015hgo,Ferretti:2016upr}, in theories with vector-like confinement~\cite{Kilic:2009mi} or in models of EW baryogenesis \cite{Jeong:2018ucz}. 

The first observation of this paper is that many existing search strategies for light resonances in the $2-20$~GeV range~\cite{ALEPH:2005ab,Adriani:1992zm,Lees:2011wb,Acciarri:1994gb,Knapen:2016moh,Izaguirre:2016dfi,Dolan:2017osp} lose sensitivity as soon as the gluon coupling is switched on.
The main reason is that the decay width into gluons dominates over the one into photons unless a non-generic hierarchy of couplings is assumed, therefore strongly suppressing  the signals expected in the existing strategies.

The dominant di-jet final states are much more difficult to distinguish from the SM background than diphotons.\footnote{As an example the LEP limit on $\text{BR}(Z\to \gamma a)$ is $1.7\cdot 10^{-5}$ from $36.9\text{ pb}^{-1}$ of data if $a$ is a diphoton resonance \cite{Abreu:1994du} and $4.7\cdot 10^{-4}$ from $5.5\text{ pb}^{-1}$ of data if $a$ is a dijet resonance \cite{Adeva:1991dw}.}  
As a way to overcome this issue, we show that the large production rate in $pp$ collisions induced by the non-zero gluon coupling can be exploited at LHCb, which already has a low mass diphoton trigger designed to look for the rare decay $B_s \to \gamma\gamma$. To substantiate this point, we use 80~pb$^{-1}$ of public LHCb diphoton data~\cite{Benson:2314368} around the $B_s$ mass to derive a limit of $\mathcal{O}(100)$~pb on the signal strength of new diphoton resonances. 
This limit already constitutes the strongest existing probe for ALPs in the mass range between 4.9 and 6.3 GeV and motivates a dedicated LHCb search for diphoton resonances in a broader mass range. We estimate the sensitivity of such a search and show that decay constants at around the TeV scale are within reach of the high-luminosity phase of LHCb. This extends the coverage of low-mass resonance searches down to masses as low as 2~GeV and constitutes a new probe of multi-TeV scale NP which could be difficult to produce directly at the LHC. 
A similar point was made in Ref.~\cite{Mariotti:2017vtv} with ATLAS, CMS, and Tevatron diphoton searches, that are however limited by trigger issues to masses roughly above 10 GeV.

We finally discuss bounds on light resonances produced from SM meson decays. We estimate the BABAR constraint on $\Upsilon(1,2,3S) \to \gamma a (jj)$ and assess the future Belle-II sensitivity. This production channel currently constitutes the best probe of ALPs below $\sim$ 3~GeV.

\section{Results}
\label{sec:ALPs}

\begin{figure}[t]
\includegraphics[width=0.49\textwidth]{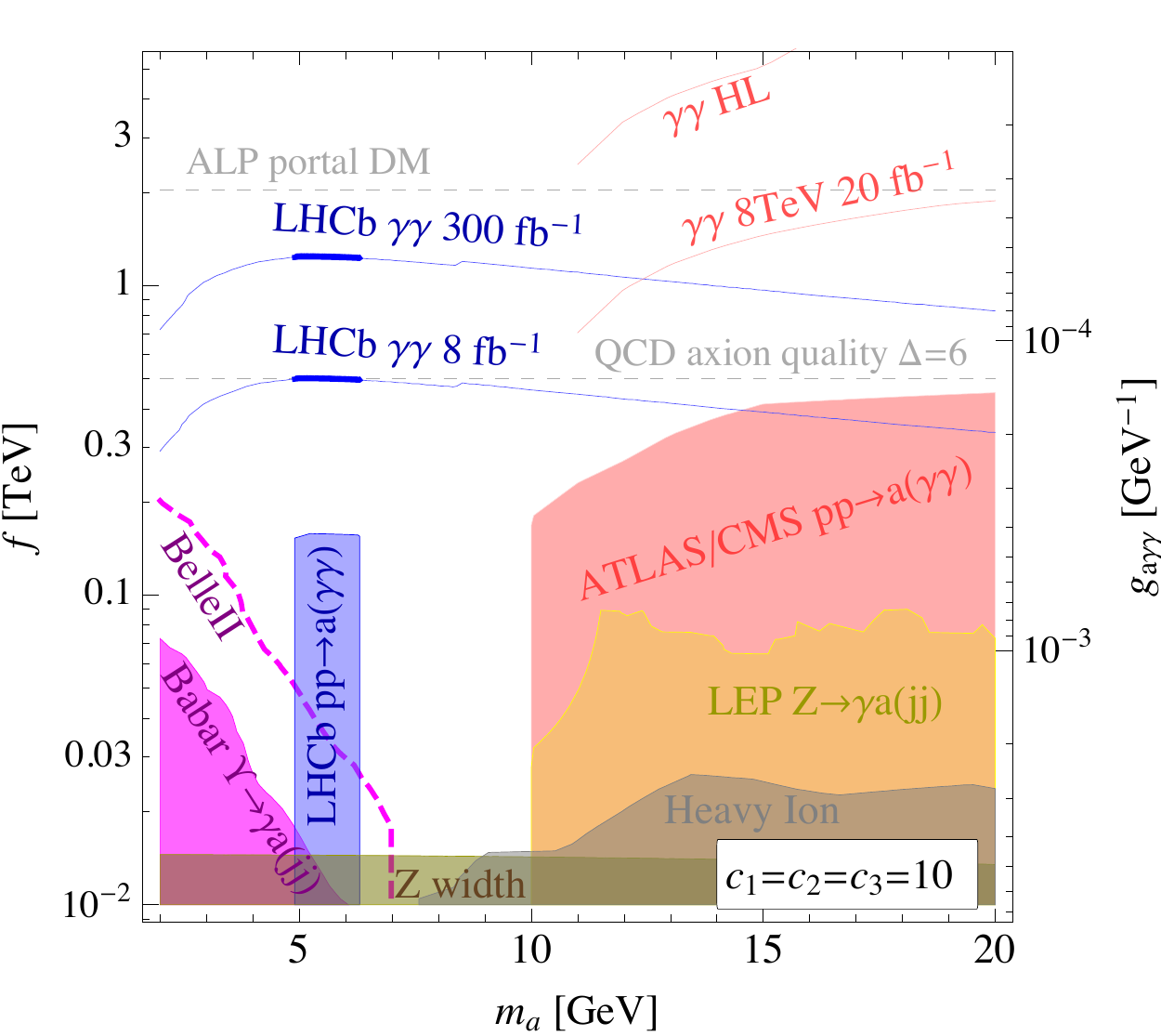}
\caption{\label{fig:ALP} Limits (shaded regions) and sensitivities (colored lines) on the ALP parameter space described in Eq.~(\ref{eq:Lagrangian}). The bounds from Babar and LHCb are first derived here from data in ~\cite{Lees:2011wb,Benson:2314368}, projections are given for Belle II and future LHC stages. Details are given in Sec.~\ref{sec:bound}.
The other bounds are derived from Z width measurements~\cite{ALEPH:2005ab,deBlas:2016ojx}, heavy ion collisions~\cite{Aaboud:2017bwk,Sirunyan:2018fhl}, $Z\to\gamma a(jj)$ decays at LEP I~\cite{Adriani:1992zm} and diphoton cross section measurements at CDF (relevant only for $m_a \simeq 10$~GeV), CMS and ATLAS~\cite{Mariotti:2017vtv}. For the latter we also give sensitivities up to the HL stage as derived in~Ref.~\cite{Mariotti:2017vtv}. The thin dashed lines indicate theory benchmarks motivated by heavy QCD axion models and by ALP-portal Dark Matter described in Sec.~\ref{sec:theory}. New coloured and EW states are expected to have masses of order $g_*f$, where $g_* =  4\pi/\sqrt{N_\text{mess}} = 4\pi/\sqrt{2\,c_i}$.
}
\end{figure}

\medskip

%Following the logic explained above, 
We consider a spontaneously broken approximate $U(1)$ symmetry in the UV.
Integrating out the new physics sector at the scale $M_\mathsmaller{\rm NP}$, we write down the effective interactions between the pNGBs and the SM
\beq
\mathcal{L}_{\rm eff} = \frac{1}{2}(\partial_\mu a)^2 - \frac{1}{2}m_a^2 a^2 + \frac{a}{f}\sum_{i=1}^3 c_i\frac{\alpha_i}{4\pi}\,F_{i,\mu\nu}\tilde{F}^{\mu\nu}_i,
\label{eq:Lagrangian}
\eeq
where $i$ runs over the hypercharge, weak and strong gauge groups, $\tilde{F}^{\mu\nu}_i = \epsilon^{\mu\nu\rho\sigma}F_{i,\rho\sigma}/2$,
$\alpha_i = g_i^2/4\pi$ and $\alpha_1$ is GUT-normalised ($\alpha_1 = 5\alpha_y/3$). The constants $c_i$ are anomaly coefficients which depend on the number of degrees of freedom chiral under the $U(1)$ symmetry and carrying a non-zero charge under the SM gauge group.\footnote{If the SM fermions and the Higgs doublet are uncharged under the $U(1)$ symmetry, the couplings of the pNGB to them arise only from loops of SM gauge bosons and can safely be neglected.}
%\FS{for the phenomenology of interest for this paper.}.

In the NP sector, the strength of the interaction $g_\ast$ generically limits the maximal number of degrees of freedom to be below $\approx (4\pi)^2/g_\ast^2$. Therefore, a lower $g_\ast$ allows for large couplings of the ALP to the SM but at the same time it lowers the scale of new physics $M_\mathsmaller{\rm NP} \simeq g_\ast f$.

For $m_a\lesssim M_Z$, we can write the ALP couplings to photons and gluons below EWSB using the same notation of the QCD axion 
\begin{equation}
\mathcal{L}_{\rm eff}\supset\frac{N\alpha_3 }{4\pi}\frac{a}{f}G_{\mu\nu}\tilde{G}^{\mu\nu}+\frac{E\alpha_{\text{em}}}{4\pi}\frac{a}{f} F_{\mu\nu}\tilde{F}^{\mu\nu}\, ,\label{eq:EFTaxion}
\end{equation}
where we have
\begin{equation}
N=c_3\quad ,\quad E = c_2 + 5 c_1/3 \quad ,\quad g_{a\gamma\gamma}=\frac{\alpha_{\text{em}}}{\pi f}\,E\, ,\label{def:axion}
\end{equation}
where $g_{a\gamma\gamma}$ agrees with the standard formula for the QCD axion after normalizing the decay constant with respect to the QCD coupling $f= 2 N f_\text{PQ}$. 
The relevant decay widths of the pNGB are
\beq
\Gamma_{\gamma\gamma} = \frac{\alpha_{\text{em}}^2 E^2}{64\pi^3}  \frac{m_a^3}{f^2},\qquad
\Gamma_{gg} = K_{gg}\frac{\alpha_s^2N^2}{8\pi^3}  \frac{m_a^3}{f^2},
\label{eq:widths}\eeq
where we include NNLO corrections to the gluon width~\cite{Chetyrkin:1998mw} in $K_{gg}$ (see Appendix~\ref{app:signal} for more details).
Note that $ (0.1~{\rm mm})^{-1}\ll \Gamma_{\rm tot} = \Gamma_{gg} + \Gamma_{\gamma\gamma } \ll m_{\gamma\gamma}^{\rm bin}$ over the mass range of our interest. The new resonance decays promptly and has a very narrow width compared to its mass.  

\medskip

The LHCb constraint and sensitivities derived in Section~\ref{sec:bound} are displayed on the ALP parameter space in Figure~\ref{fig:ALP}, for the benchmark $c_1 = c_2 = c_3 = 10$.
We compute $\sigma(pp \to a)$ with ggHiggs v4~\cite{Ball:2013bra,Bonvini:2014jma,Bonvini:2016frm,Ahmed:2016otz} using the {\tt mstw2008nnlo} pdf set.
We compare it with that obtained by the use of different pdf sets and of MadgraphLO\_v2\_6~\cite{Alwall:2011uj,Alwall:2014hca} upon implementing the ALP model in FeynRules~\cite{Alloul:2013bka}, finding differences from $~20\%$ at $m_a = 20$~GeV to a factor of 2 or larger for $m_a < 5$~GeV. As detailed in Appendix~\ref{app:signal}, a more precise determination of the signal would be needed, especially for $m_a \lesssim 5~$GeV.
%We compute $\sigma(pp \to a)$ with MadgraphLO %plus Pythia plus matching up to an extra jet, upon implementing the ALP model in FeynRules~\cite{Alloul:2013bka}, and using the {\tt cteqsl1} pdf set.
%For the range of $m_a$ of our interest, we compare $\sigma(pp \to a)$ with that obtained by the use of ggHiggs v4~\cite{Ball:2013bra,Bonvini:2014jma,Bonvini:2016frm,Ahmed:2016otz} and of different pdf sets, finding differences from $~20\%$ at $m_a = 20$~GeV to a factor of 2 or larger for $m_a < 5$~GeV.
%This underlines the need of . We refer the reader to Appendix~\ref{app:signal} for more quantitative details.

In Figure~\ref{fig:ALP} we also show
\begin{itemize}

\item[i)] the 2$\sigma$ constraint $\Gamma_Z - \Gamma_Z^{\rm SM} < 5.8$~MeV~\cite{ALEPH:2005ab,deBlas:2016ojx};
\item[ii)] the LEP limit BR$(Z \to \gamma a (jj)) < 1-5\times10^{-4}$~\cite{Adriani:1992zm};
\item[iii)] the constraint derived in~\cite{Mariotti:2017vtv} from the ATLAS~\cite{Aad:2012tba,Aaboud:2017vol}, CMS~\cite{Chatrchyan:2014fsa}, and CDF\cite{Aaltonen:2011vk} inclusive diphoton cross section measurements, corresponding to $\sigma(pp/p\bar{p} \to X\,a(\gamma\gamma)) <10-100$~pb;
\item[iv)] the sensitivities derived in~\cite{Mariotti:2017vtv} from inclusive diphoton cross section measurements at ATLAS and CMS.
The HL-LHC reach assumes minimal photon $p_T$ cuts of 25 and 22~GeV and minimal photon separation of $\Delta R=0.4$. These numbers correspond to the 7 TeV measurement in Ref.~\cite{Aad:2012tba}.
Higher $p_T$ cuts would increase the minimal value of the invariant mass within the reach of HL-LHC.
\item[v)] the BABAR constraint BR$\big(\Upsilon_{2S,3S} \to~\gamma a(jj)\big) < 10^{-4} - 10^{-6}$~\cite{Lees:2011wb}, where we compute
 %the width for the decays $\Upsilon(2S,3S) \to \gamma a$ mediated by a virtual photon in s-channel
\beq
\frac{{\rm BR}\big(\Upsilon \to \gamma a \big)}{{\rm BR}\big(\Upsilon \to \mu \bar{\mu}\big)}
\simeq 2 E^2 \frac{\alpha_{\rm em}}{4\pi}\,\Big(\frac{m_\Upsilon}{4\pi f}\Big)^{\!2} \Big(1-\frac{m_a^2}{m_\Upsilon^2}\Big)^{\!3},
\label{eq:Upsilon}
\eeq
where ${\rm BR}\big(\Upsilon_{2S,3S} \to \mu \bar{\mu}\big)=1.92 \%, 2.18 \%$. The above expression agrees with the one of Ref.~\cite{Masso:1995tw}.  
\item[vi)] the Belle-II sensitivity in the same channel, that we determine simply by rescaling the expected sensitivities in~\cite{Lees:2011wb} by a factor of 10.
This assumes that the Belle-II reach will be statistics-dominated, and that it will be based on a factor of 100 more~$\Upsilon(3S)$ than the BABAR one (i.e. on $\simeq 1.2 \times 10^{10}~\Upsilon(3S)$ in total).
The current Belle-II run plan for the first years assumes only a factor of 10 for the above ratio~\cite{BelleII:Komarov,Kou:2018nap}, corresponding to a few weeks of dedicated run at the $\Upsilon(3S)$ threshold. An extra factor of 10 could be obtained in a comparable time with dedicated later runs, because  a higher instantaneous luminosity is foreseen~\cite{Kou:2018nap}.
An analogous search could be effectively performed, at Belle-II, also analysing the decays of $\Upsilon(1S,2S)$.
%\FS{Why did Babar use the $\Upsilon(2S,3S)$ from dedicated runs, and not those from $\Upsilon(4S,5S,\dots)$ decays?}
%
\item[vii)] limits from the diphoton final state from heavy ion collisions are extracted from the recent CMS analysis in Ref.~\cite{Sirunyan:2018fhl} and the reinterpretation of the ATLAS  light by light scattering data \cite{Aaboud:2017bwk} of Ref.~\cite{Knapen:2017ebd}. The lower reach of these measurements is set to $m_a\gtrsim 5\text{ GeV}$ as a consequence of the minimal cuts on the two photons transverse momenta. 
\end{itemize}
ATLAS limits from $Z\to\gamma a(\gamma\gamma)$~\cite{Aad:2015bua} are not displayed in Fig~\ref{fig:ALP}.
They imply $\text{BR}(Z\to\gamma a(\gamma\gamma))< 2.2\cdot10^{-6}$ and turn out to be comparable to the heavy ions bound for our benchmark in Fig.~\ref{fig:ALP}. Similar constraints can be derived from the ATLAS inclusive search in $p p\to \gamma a(\gamma\gamma)$~\cite{Aad:2015bua}. The lower invariant mass reach of these ATLAS searches is set by the diphoton isolation requirement of~\cite{Aad:2015bua}, $\Delta R_{\gamma\gamma}=0.15$. This  corresponds to an ALP mass of 4 GeV as discussed in Ref.~\cite{Jaeckel:2015jla}.
Notice that LEP searches for~$Z\to\gamma a(\gamma\gamma)$~\cite{Acciarri:1994gb} are weaker than the ATLAS bound. Future sensitivities from $e^+e^-\to \gamma a(\gamma\gamma)$~\cite{Izaguirre:2016dfi,Dolan:2017osp} do not reach values of $f$ larger than $\simeq$~50~GeV and are not shown. Finally, the proposed search in $B \to K^{(*)} a(\gamma\gamma))$~\cite{Izaguirre:2016dfi} at Belle-II has some sensitivity in a very limited portion of our mass range and it is not shown to avoid clutter.  

\begin{figure}[h]
\includegraphics[width=0.49\textwidth]{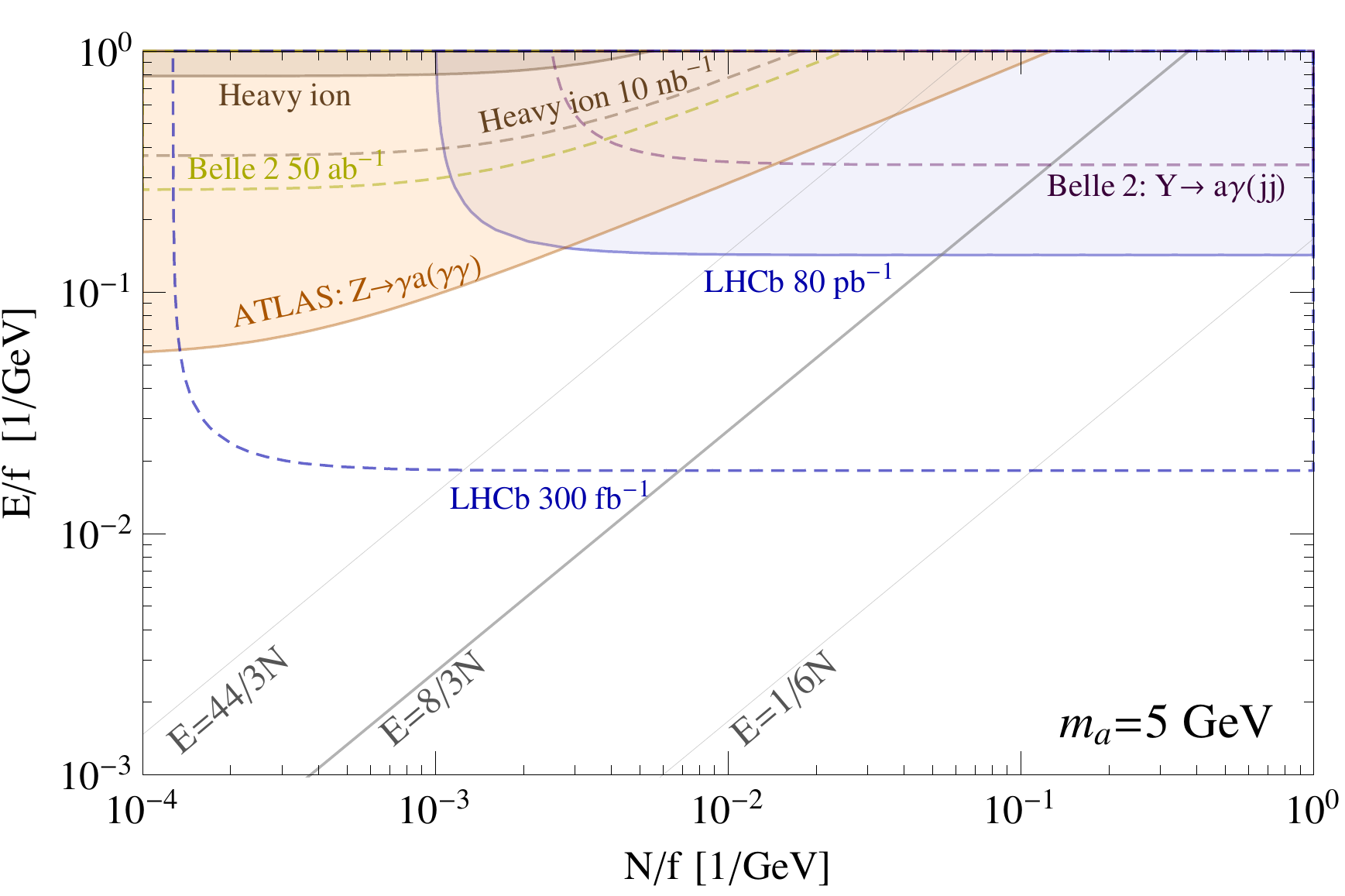}\hfill
\includegraphics[width=0.49\textwidth]{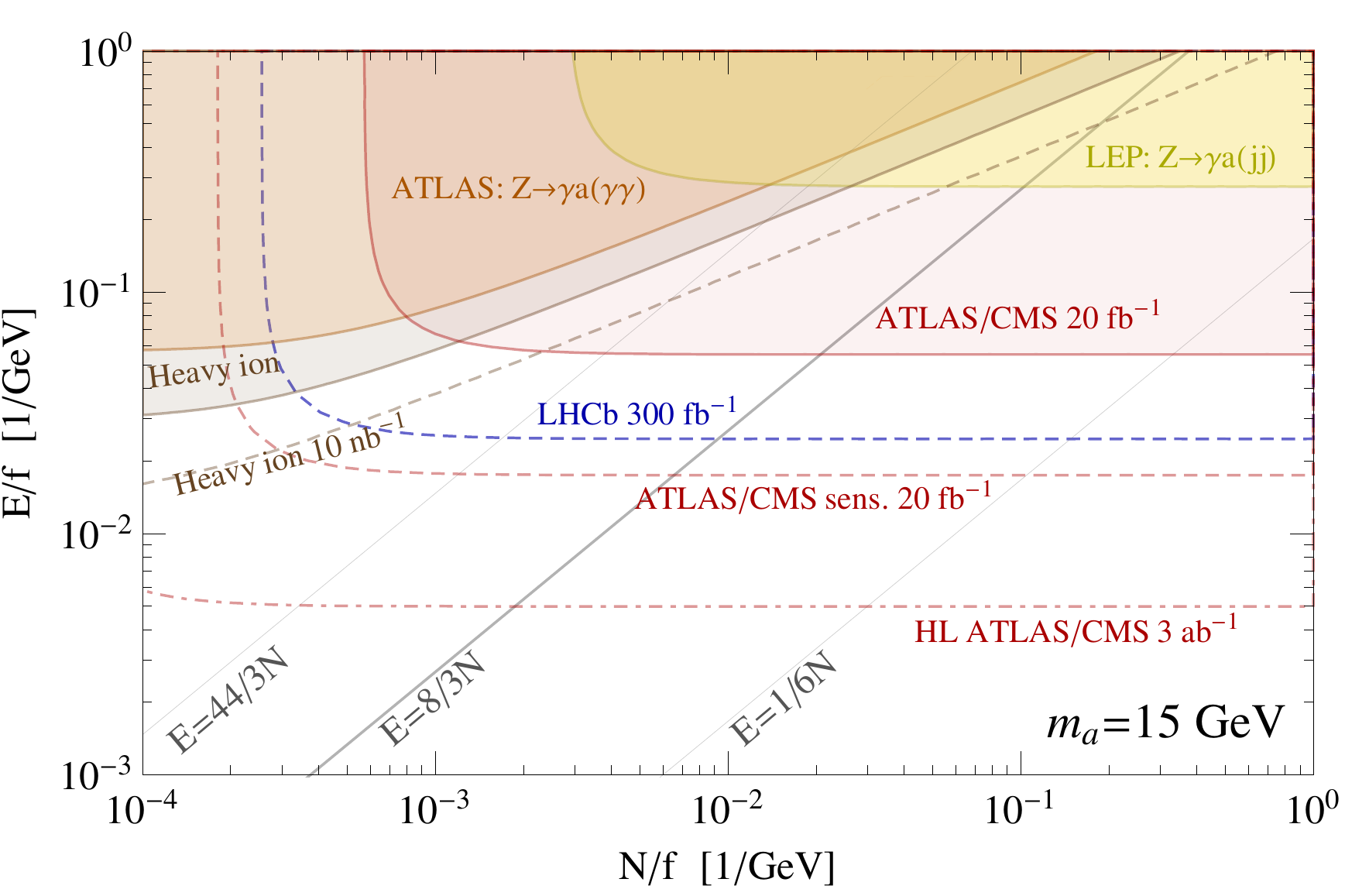}
\caption{\label{fig:ALP2} Constraints on the ALP parameter space for fixed masses $m_a=5,15\text{ GeV}$ in the up, down panel respectively.
We fix $c_1=c_2$ so that $E$ in Eq.~\eqref{def:axion} controls both the $Z\gamma$ and the $\gamma\gamma$ coupling.
The bounds are shown as shaded regions while the projections as dashed lines. The three grey lines show the ``axion window'' obtained by integrating out fermions in different representations of the SM gauge group, the central one $E=8N/3$ corresponds to the choice of Figure~\ref{fig:ALP}.} 
\end{figure}

In Fig.~\ref{fig:ALP2} we fix the ALP masses to two representative values $m_a=5,15\text{ GeV}$ and show the impact of the various searches in the plane $(N/f, E/f)$ which control the ALP's gluon and photon coupling respectively.  As one can see from Fig.~\ref{fig:ALP2}, diphoton searches for a ALP produced in gluon fusion both at ATLAS/CMS (see Ref. \cite{Mariotti:2017vtv}) and at LHCb (see Sec.~\ref{sec:bound}) can be sensitive to $N/f$ as small as $10^{-4}\text{ GeV}^{-1}$  as long as the coupling to the photons is large enough. Moreover they can cover significant portion of the parameter space where the couplings are of their natural size.

Searches taking advantage of uniquely the photon coupling such as the ones in Refs.~\cite{Knapen:2016moh,Dolan:2017osp,Aad:2015bua} become relevant only in the upper left corner of the plane where $E/N\gtrsim 50$. Such a hierarchy can be realized in clockwork constructions where the photon coupling is enhanced with respect to the gluon one (see for example Ref.~\cite{Farina:2016tgd}). 

The ATLAS, CMS and LHCb limits and sensitivites shown in Fig.~\ref{fig:ALP2} are derived assuming gluon fusion as the ALP production process, so they sharply stop at a given small gluon coupling. 
If other production processes like vector-boson-fusion are taken into account, the limits and sensitivities would be slightly improved in the upper left corner of Fig.~\ref{fig:ALP2}. Practically, the Heavy Ion results that we are including will always lead to stronger constraint because of the enhanced photon-fusion production and the loop suppressed background from light-by-light scattering.

The bottom right corner where the new resonance mostly couples to gluons is challenging to constrain in this mass range, even though boosted dijet searches at the LHC were recently able to go down to invariant masses of 50 GeV (see Refs.~\cite{Mariotti:2017vtv, Sirunyan:2017nvi, Arganda:2018cuz}). Of course for $N/f\gtrsim (100 \rm GeV)^{-1}$ one expects color states generating the ALP coupling to be within the reach of the LHC.   

%Estimating the the sensitivity of Belle-II in $e^+e^-\to \gamma a(jj)$ and $B \to K^{(*)} a(jj)$ goes beyond the purposes of this paper.
%The reason why the latters bounds with  is the extremely small BR$(a\to\gamma\gamma)$, which is a generic feature every time the QCD coupling is switched on.

%Higgs decays to $aa$ are also totally negligible given the Lagrangian in Eq.~(\ref{eq:Lagrangian}), therefore related searches (like~\cite{Khachatryan:2016vau,Aad:2015oqa}) do not give constraints in our parameter space.
%They would constrain values of $f$ of few hundreds of GeV if 
%Note that the ALP couplings to fermions generated by those in the Lagrangian~(\ref{eq:Lagrangian}) are suppressed as...blabla

%Previous ALPs at colliders~\cite{Mimasu:2014nea,Jaeckel:2015jla,Bauer:2017ris}
 
\section{Physics cases}
\label{sec:theory}
In this section we expand on the two theory lines displayed in Fig.~\ref{fig:ALP}. We would like to motivate: 1) the coupling of the axion to gluons and photons, 2) the TeV decay constant, 3) the mass range considered here.

\paragraph{\bf Heavy axions} As a first example, we consider a particular class of axion solutions to the strong CP problem in the SM. First of all, introducing a spontaneously broken Peccei-Quinn symmetry $U(1)_{\text{PQ}}$ which is anomalous under QCD~\cite{Peccei:1977ur,Peccei:1977hh} leads unavoidably to a light axion with non-zero couplings to gluons~\cite{Weinberg:1977ma,Wilczek:1977pj}. In this sense, the axion coupling to gluons is deeply connected to its role in solving the strong CP problem. Taking the SM fields to be uncharged under the $U(1)_{\text{PQ}}$, the QCD anomaly is generated by heavy vector-like fermions like in KSVZ type of models \cite{Kim:1979if,Shifman:1979if}
\begin{equation} 
\mathcal{L}_{\text{PQ}}\supset g_{\ast}\Phi \psi \tilde{\psi} +\text{h.c.},\quad   \Phi= \frac{f}{\sqrt{2}} e^{ia/f}\ ,
\label{eq:KSVZ}
\end{equation}
where the fermion charges should satisfy $\vert q_\psi^{\text{PQ}}-q_{\tilde{\psi}}^{\text{PQ}}\vert=q_\Phi^{\text{PQ}}$ and by writing Eq.~\eqref{eq:KSVZ} we take $q_\Phi^{\text{PQ}}=1$.  After $U(1)_{\text{PQ}}$ gets spontaneously broken by the VEV of $\Phi$, the fermion mass is at $M_\mathsmaller{\rm NP}=g_\ast f/\sqrt{2}$. Below the PQ breaking scale we can integrate out the heavy fermions and match to the effective Lagrangian in Eq.~\eqref{eq:EFTaxion}:
\begin{equation}
N=q_\Phi^{\text{PQ}}\sum_{\psi} C_3(R_\psi)\qquad E=q_\Phi^{\text{PQ}}\sum_{\psi}Q_{\text{em}}^2(R_\psi)\ .
\end{equation}
The vector-like fermions are often assumed to carry a non-zero hypercharge in order to allow a non-zero mixing with the SM quarks, to make them decay avoiding cosmological problems.
%Typically, the vector-like fermions should carry a non-zero hypercharge in order to allow a non-zero mixing with the SM quarks to make them decay avoiding cosmological problems.
This induces an anomaly of $U(1)_{\text{PQ}}$ with respect to the hypercharge, which leads to a non-zero coupling of the axion to photons: $E\neq 0$.
To fix a benchmark, we add $N_{\text{mess}}$ complete $SU(5)$ fundamental representations,
that lead to $N=N_{\text{mess}}/2$ and $E=4/3 \,N_{\text{mess}}$.
This is the scaling assumed in Fig.~\ref{fig:ALP}, where we also take $N_{\text{mess}}=(4\pi/g_{\ast})^2$ to ensure calculability below $M_\mathsmaller{\text{NP}}$. In Fig.~\ref{fig:ALP2} we go beyond this benchmark and show how E/N can be modified changing the SM representation of the fermions in Eq.~\eqref{eq:KSVZ} (see Ref.~\cite{DiLuzio:2016sbl} for a discussion). 

%\FS{[For Diego: the line in the plot has $\Delta = 6$ and $\Lambda_\text{UV} = 10^{15}$~GeV]}

Operators breaking $U(1)_{\text{PQ}}$ other than the QCD anomaly would in general spoil the axion solution of the strong CP problem~\cite{Holman:1992us,Kamionkowski:1992mf,Barr:1992qq,Ghigna:1992iv}.
We can parametrize these contributions as new terms in the potential for the scalar~$\Phi$:  
\begin{equation}
\Delta V_{\slashed{\text{PQ}}}=\lambda_\Delta\frac{\Phi^\Delta}{\Lambda_{\text{UV}}^{\Delta-4}}+\text{h.c.},
\qquad \lambda_\Delta=\vert \lambda_\Delta\vert e^{i\alpha_\Delta}\ .
\label{eq:PQbreaking}
\end{equation}
In the presence of these new contributions the axion potential below the QCD phase transition is
\begin{equation}
V_{a}\simeq - \Lambda_\text{QCD}^4\cos\frac{N a}{f}
+\frac{1}{2^{\frac{\Delta}{2}-1}}\frac{\vert \lambda_\Delta\vert f^{\Delta}}{\Lambda_{\text{UV}}^{\Delta-4}}\cos\Big(\alpha_{\Delta}+ \Delta \frac{a}{f}\Big)\ .
\end{equation} 
Since the new phase $\alpha_\Delta$ is in general not aligned with the contribution given by the QCD anomaly, the presence of the UV operator shifts the axion VEV away from the origin, jeopardizing the solution to the strong CP problem. Note that this holds even if the NP sector inducing Eq.~(\ref{eq:PQbreaking}) preserves CP, because a new phase $\alpha_\Delta \sim O(1)$ is induced by rotating away the phase in the quark mass matrices.
 Requiring $2\langle N a/f\rangle\lesssim 10^{-10}$  to satisfy the present bound on the neutron dipole moment~\cite{Afach:2015sja,Pospelov:2005pr} gives an upper bound on the axion decay constant  
\begin{equation}
f\lesssim\Lambda_{\text{UV}}\left[10^{-10} \cdot\frac{N}{\Delta}\cdot\left(\frac{\Lambda_{\text{QCD}}}{\Lambda_{\text{UV}}}\right)^4 \right]^{1/\Delta}\,,\label{eq:quality}
\end{equation}
where we have assumed~$\vert \lambda_\Delta\vert \sim\alpha_{\Delta}\sim\mathcal{O}(1)$ and neglected other $O(1)$ factors for simplicity.
The upper bound on $f$ depends on the scale of the UV completion $\Lambda_{UV}\gtrsim g_{\ast} f$ and on the ``quality'' of the $U(1)_{\text{PQ}}$, i.e. the dimension $\Delta$ of the lowest dimension operator breaking the symmetry.
 
In the best case scenario, first discussed in Refs~\cite{Kamionkowski:1992mf,Holman:1992us,Barr:1992qq,Ghigna:1992iv}, the $U(1)_{\text{PQ}}$ is only broken by Planck suppressed operators\footnote{Gravity is expected to break global symmetries at the non perturbative level via wormhole solutions swallowing the PQ charge~\cite{Giddings:1987cg}. In this case the Wilson coefficient of the operators in Eq.~\eqref{eq:PQbreaking} can be very suppressed for a large enough wormhole action: $\vert \lambda_\Delta\vert\sim e^{-S_{\text{Eucl}}}$.
The latter has been shown in Ref.~\cite{Kallosh:1995hi} to be too small in the Einstein theory of gravity but large enough in theories where the Einstein theory is suitably modified at Planckian distances.} but, more generally, one might argue that all the global symmetries should be an accidental consequence of the gauge and matter content of the theory, exactly like in the SM. In the latter case the $\Lambda_{\text{UV}}$ in Eq.~\eqref{eq:quality} will be below $M_{\text{Pl}}$.
Taking Eq.~\eqref{eq:quality} at face value, the most dangerous contribution comes from $\Delta=5$ operators, that would require $f \lesssim O(10)~\text{ GeV}$ even for $\Lambda_\text{UV} =M_{\text{Pl}}$. However, if operators of dimension five are forbidden (for example by a discrete $Z_2$-symmetry) then $\Delta=6$ contributions give $f \lesssim O(10) \text{ TeV}$ for $\Lambda_\text{UV} =M_{\text{Pl}}$ and $f \lesssim O(1) \text{ TeV}$ for $\Lambda_\text{UV} =M_{\text{GUT}}$, motivating the ranges of decay constant of interest for this paper. Having $f$ around the TeV scale would lead to axion solutions relying on $U(1)_{\text{PQ}}$ with the same quality of the baryon number in the SM. 

In the usual QCD axion where $m_a \simeq 6 \text{ keV}\cdot \text{TeV}/f$ (see e.g.~\cite{diCortona:2015ldu}), values of the decay constant motivated by the axion quality problem are abundantly excluded by star cooling bounds \cite{Raffelt:1996wa} and $K\to \pi a$ transitions~\cite{Georgi:1986df,Freytsis:2009ct}. A common solution to this problem is to go to higher values of $f$ and require a $U(1)_{\text{PQ}}$ with higher quality. Such a $U(1)_{\text{PQ}}$ can be made accidental in extra-dimensions or with more complicated UV completions in 4 dimensions (we refer to Refs~\cite{Randall:1992ut,Redi:2016esr,DiLuzio:2017tjx,Duerr:2017amf} for an illustration of the challenges involved in constructing gauge theories with a $U(1)_{\text{PQ}}$ with arbitrarily high quality). 

Alternatively, one can construct QCD axion models where the axion mass is heavier than its QCD value. The idea is to introduce new contributions to the axion potential which are aligned to the QCD one, so that the axion mass gets larger without spoiling the solution to the strong CP problem.
A larger $m_a$ then relaxes the experimental constraints on $f$, potentially allowing to satisfy Eq.~(\ref{eq:quality}).
There are several classes of models of this type which differ from the way the alignment is achieved: mirror axion models with one axion and two mirror QCD's \cite{Rubakov:1997vp,Berezhiani:2000gh,Hook:2014cda,Fukuda:2015ana,Dimopoulos:2016lvn}, models where the QCD running is modified at high energies \cite{Holdom:1982ex,Choi:1988sy,Holdom:1985vx,Dine:1986bg,Flynn:1987rs,Choi:1998ep}, and a more recent proposals \cite{Agrawal:2017ksf} where the QCD group is embedded in $SU(3)^N$ with $N$ axions relaxing each one of the allowed $\theta$-angles.   

All the solutions of the strong CP problem mentioned above can easily achieve the 2-20 GeV mass range, and result in an axion which generically couples to both gluons and photons with a decay constant at the TeV scale or lower. These are a perfect benchmark for the collider searches discussed here. For illustrative purposes we show in Fig.~\ref{fig:ALP}  the value of $f$ corresponding to a $U(1)_{\text{PQ}}$ broken by $\Delta=6$ operators generated at~$M_{\text{GUT}}=10^{15}\text{ GeV}$.

\paragraph{\bf ALP-mediated Dark Matter}
The second example of ALP with coupling and masses of interest for this study comes from demanding it to be the mediator that couples the SM to fermion DM, singlet under the SM gauge group.
This possibility has particular interest for colliders because direct detection constraints are totally irrelevant, see e.g.~\cite{Bishara:2017pfq}.

We write the ALP coupling to DM as in equation~(\ref{eq:KSVZ}) and identify the DM as the Dirac fermion $(\psi,\tilde \psi^\dagger)$, so that $m_\psi = g_\ast f/\sqrt{2}$.
The DM annihilation cross section into SM particles, mediated by the ALP, is dominated by final state gluon pairs and reads 
\beq
(\sigma v)_{gg} = \frac{2}{\pi}\,\Big(\frac{c_3\,\alpha_s}{4 \pi}\Big)^2 \frac{g_\ast^2}{f^2}\,,
\label{eq:sigmav_gluons}
\eeq
where $\alpha_s$ is evaluated at the scale $\mu = 2 \,m_\psi$.
The cross section for $t$-channel annihilation into a pair of mediators is $p$-wave and reads~\cite{DEramo:2016aee}
\beq
(\sigma v_\text{rel})_{aa} = \frac{v_\text{rel}^2}{384\pi} \frac{g_\ast^4}{m_\psi^2}\,,
\label{eq:sigmav_ALPs}
\eeq
therefore it is negligible with respect to the annihilation into gluons for the parameter values we are interested in, even for relativistic $v_\text{rel}$.
Requiring Eq.~(\ref{eq:sigmav_gluons}) to match $4.8 \times 10^{-26}$ cm$^3$/sec, which is the value needed for heavy Dirac DM to reproduce the correct DM abundance via thermal freeze-out~\cite{Steigman:2012nb}, we find
\beq
m_\psi \simeq 4.6~\text{TeV} \frac{c_3}{10} \, \Big(\frac{g_\ast}{3}\Big)^2 \; \Rightarrow \; f \simeq 1.9~\text{TeV} \,\frac{3}{g_\ast}\,,
\label{eq:sigmav_ALPs}
\eeq
where in the second equality we have assumed the scaling $c_3 \simeq 8\pi^2/g_\ast^2$.
This is the benchmark value we display in Figure~\ref{fig:ALP}. It is interesting to note that indirect detection is still far from probing thermal values of the annihilation cross section for DM in this mass range (see e.g.~\cite{Ackermann:2015zua,Giesen:2015ufa,Abdallah:2016ygi}), thus adding further motivation to test this scenario with colliders.

Note that we have neglected the possible Sommerfeld enhancement from exchange of the ALP in the initial state.
The precise computation of this effect is still the object of some debate, see e.g.~\cite{Bellazzini:2013foa} for a recent study with references, so that for simplicity we do not include it here.
Its inclusion would result in an $O(1)$ change in the favoured value of $f$, but would not affect our physics point that pseudoscalar mediated DM motivates ALP searches at flavor factories.

\paragraph{\bf R-axion in Supersymmetry.}
We finally notice that the simplified DM model presented above arises naturally in theories of low-scale SUSY breaking. These predict that the lightest supersymmetric particle (LSP) is the Gravitino, whose mass $m_{3/2}$ is generically too small to account for the observed DM abundance.
Indeed, using the power counting  described in~\cite{Bellazzini:2017neg}, one gets $m_{3/2} = F/(\sqrt{3} M_\text{Pl}) \simeq 11~\text{meV}\cdot (g_\ast/3)\cdot(f/4~\text{TeV})^2$. While not reproducing the observed value of DM, Gravitino masses in this ballpark are safe both from collider~\cite{Brignole:1997sk,Brignole:1998me,Maltoni:2015twa} and cosmological~\cite{Osato:2016ixc} constraints.

In the absence of stable superpartners, the natural DM candidate in these SUSY theories are particles belonging to the messenger or SUSY breaking sectors, see~\cite{Dimopoulos:1996gy} for a first study of this possibility. In this case, as first noted in~\cite{Mardon:2009gw} (see~\cite{Fan:2010is} for further model building), the DM phenomenology may be dominated by its interactions with a pseudoscalar that is naturally present in the theory, the $R$-axion.

This arises as the pNGB of the $U(1)_R$ symmetry, defined as the only abelian global symmetry which does not commute with the SUSY generators.
The spontaneous breaking of the $U(1)_R$ is intimately related to SUSY-breaking according to the general results of~\cite{Nelson:1993nf,Intriligator:2007py}.
The $R$-axion couplings to gluons and photons are unavoidably generated by loops of gauginos, whose Majorana masses are chiral under the $U(1)_R$, and possibly by UV messengers. Couplings to fermions and to the Higgs are less generic and can be suppressed by suitable charge assignment (see~\cite{Bellazzini:2017neg} for more details). Under these circumstances, the R-axion matches perfectly the Lagrangian in Eq.~(\ref{eq:Lagrangian}).

For $f = O$(TeV), motivated here not only by DM but also by the naturalness of the Fermi scale, i) its mass is expected to lie in the MeV range~\cite{Bagger:1994hh} or above~\cite{Intriligator:2007py,Bellazzini:2017neg}, thus motivating searches at flavor factories, ii) superpartners can be taken outside the LHC reach, thus making it potentially the first sign of SUSY at colliders~\cite{Bellazzini:2017neg}.

\section{Diphoton Searches at ${\rm LHCb}$}
\label{sec:bound}
%\FS{[I wrote this section in a way that one just needs to look at the equations and the figure to understand (almost) everything]}
LHCb detects photons %are detected either from their energy deposition in the ECAL (``unconverted'' photons), or from the energy deposition and trajectory of the $e^+e^-$ pair to which they convert upon interacting with the detector material (``converted'' photons).
either as ``unconverted'', i.e. they reach the electromagnetic calorimeter (ECAL), or as ``converted'', i.e. they convert to an $e^+e^-$ pair upon interacting with the detector material before reaching the ECAL.
 The public LHCb note~\cite{Benson:2314368} presents the trigger and cut strategy that will be used to look for $B_s \to \gamma\gamma$, and classifies diphoton events into two unconverted (0CV), one unconverted and one converted (1CV LL and DD, corresponding to conversions occurring in the Vertex Locator region or after it) and two converted (2CV) samples.
 %It also provide histograms of measured diphoton events that pass such cuts in each conversion category, and the efficiency of all cuts on the $B_s \to \gamma\gamma$ signal that makes it within the detector acceptance.

Searches for $B_s \to \gamma\gamma$ benefit from requiring the $\gamma\gamma$ vertex to be displaced from the $pp$ interaction point, while the resonances we are interested in typically have a lifetime much shorter than the $B_s$ one.
A displaced $\gamma\gamma$ vertex is however not imposed on the 0CV sample, because the resolution on the directions of the photons does not allow for a precise enough vertex reconstruction. Therefore this sample can be used to derive a bound on prompt diphoton resonances. 

Measured diphoton events that pass the cuts are reported in~\cite{Benson:2314368} for $L =$~80~pb$^{-1}$ of data, for each conversion category, in a diphoton invariant mass interval $4.9~{\rm GeV} < m_{\gamma\gamma} < 6.3~{\rm GeV}$ and in bins of 14.5~MeV.
No known QCD or SM resonance is expected to give a signal within the LHCb reach, explaining why the event distributions in $m_{\gamma\gamma}$ are very smooth in all categories, so that they constitute an ideal avenue to look for BSM resonances.
Therefore, we place an upper limit on the signal cross section of a resonance $a$ decaying to diphotons as
\beq
N_{\rm sig}(m_a) < 2 \sqrt{N_{\rm bkg}\,\frac{m_{\gamma\gamma}^{\rm bin}}{14.5~{\rm MeV}}}\, ,
\label{eq:LHCblimit}
\eeq
where
\beq
N_{\rm sig} = \epsilon \times \sigma_{\rm fid} \times L, \quad \sigma_{\rm fid} = A\times\sigma(pp \to Xa(\gamma\gamma))\, ,
\eeq
with $A$ the geometrical acceptance of the signal in the LHCb detector and $\epsilon$ the total efficiency of the cuts plus detector effects in a given diphoton category. We use
\beq
A = 0.15\, ,\qquad \epsilon_{\rm 0CV} = 0.142\, ,\label{eq:bench}
\eeq
where the latter is given in~\cite{Benson:2314368} for the SM ``signal'' $B_s \to \gamma\gamma$, and we determine the former by simulating the signal
(see Appendix \ref{app:simulations} for details)
%\footnote{We implement the ALP model of Eq.~(\ref{eq:Lagrangian}) in FeynRules~\cite{Alloul:2013bka}, we generate events with MadgraphLO\_v2\_6~\cite{Alwall:2011uj,Alwall:2014hca} and shower them with Pythia~8.1~\cite{Sjostrand:2006za,Sjostrand:2007gs}, matching up to 2 extra jets~\cite{Alwall:2007fs}. \FS{Check these refs}
%}
and imposing $2< \eta < 5$ at truth level.
%\FS{Should we verify with some simulation that for $m_a = m_{B_s}$ we obtain roughly the same cut efficiencies given in the Tables of~\cite{Benson:2314368}?}

Coming to the right-hand side of Eq.~(\ref{eq:LHCblimit}), $N_{\rm bkg}$ is the number of background events in the $14.5~{\rm MeV}$ bin reported in~\cite{Benson:2314368}, which we take constant as the distribution in $m_{\gamma\gamma}$ is actually flat well within its statistical uncertainties.\footnote{While
this holds for the 1CV and 2CV categories, the distribution in the 0CV category is flat up to $m_{\gamma\gamma} \simeq 5.7$~GeV, and then drops smoothly. A possible origin of this drop is the use of $2\times2$ ECAL cells to measure the photon energy deposition at the first level of the software trigger (HLT1)~\cite{Benson:2314368}. In Appendix~\ref{app:0CVsmooth} we verified that imposing invariant mass cuts at HLT1 can cause a flat background at HLT1 to develop a dropping shape at higher level, where the invariant mass is defined using $3\times3$ cells.}
$m_{\gamma\gamma}^{\rm bin}$ is the size of the bin centered on $m_{\gamma\gamma} = m_a$ that we expect to contain most of the signal from the resonance, which we assume to be narrow. In practice we use
\beq
\label{eq:bkg}
N_{\rm bkg} = 8000\times\frac{L}{80~\text{pb}^{-1}}, \qquad m_{\gamma\gamma}^{\rm bin} = 4\delta m_{\gamma\gamma}\,,
\eeq
where $\delta m_{\gamma\gamma}$ is the invariant mass resolution for the 0CV category which can be derived from the energy resolution and the granularity of the LHCb ECAL (see Appendix~\ref{app:fake_pions}).
Fixing for definiteness $m_{\gamma\gamma}^{\rm bin}/m_{\gamma\gamma} = 13\%$, we obtain 
\begin{equation}
\sigma_{\text{fid}}^{0\text{CV}}\lesssim 106\text{ pb}\cdot\sqrt{\frac{m_a}{5\text{ GeV}}}\cdot\sqrt{\frac{80\text{ pb}^{-1}}{\text{L}}}\, .
\end{equation}
The sensitivities that could be achieved by the current full dataset of $\simeq 8$~fb$^{-1}$ and by the High Luminosity phase of LHCb with $\simeq 300$~fb$^{-1}$ of data can be easily obtained from the above equation.\footnote{Actually only $\simeq 2$~fb$^{-1}$ have been recorded outside the $B_s$ mass window, we neglect this drop in luminosity for simplicity.}

We also extend the mass range of the search to $3 < m_{\gamma\gamma}/{\rm GeV} <  20$, where the lower bound is chosen to make the computation of the signal strength reliable (see also Appendix \ref{app:signal}) and the upper bound is chosen somehow arbitrarily at 20 GeV, where the reach of the current ATLAS/CMS inclusive diphoton dataset~\cite{Mariotti:2017vtv} is already stronger than the projections of LHCb. For simplicity we take the signal acceptance and the efficiency to be constant and equal to the ones in Eq.~(\ref{eq:bench}). 
We discuss in Appendix \ref{app:simulations} the motivations for this simplified assumption.
Moreover we assume that the background  is also constant in the extended mass range and equal to the one in Eq.~(\ref{eq:bkg}).
This simple procedure sets a useful benchmark for the actual search, which is good enough for the purpose of this paper. The resulting reach in the ALP parameter space is shown in Fig.~\ref{fig:ALP}.

We finally speculate about the limit and reach obtainable if the 1CV photon categories could be used. To set an optimistic reach, we do not take into account the signal loss because of the requirement of vertex displacement in present LHCb search. With this assumption, we repeat the procedure described above, with constant background $N_{\rm bkg}^{\rm 1CV,DD} = 1600$ and $N_{\rm bkg}^{\rm 1CV,LL} = 1300$ and constant efficiencies $\epsilon_{\rm 1CV,DD} = 1.35\%$ and $\epsilon_{\rm 1CV,LL} = 1.32\%$ as reported in Ref.~\cite{Benson:2314368}.
Concerning the mass resolution, we take the one of the 0CV category divided by $\sqrt{2}$ to roughly account for the much better energy resolution of the converted photon. With all these assumptions we combine in quadrature the exclusions from the LL and DD single-converted categories and get
\begin{equation}
\sigma_{\text{fid}}^{1\text{CV}}\lesssim  283\text{ pb}\cdot\sqrt{\frac{m_a}{5\text{ GeV}}}\cdot\sqrt{\frac{80\text{ pb}^{-1}}{\text{L}}}\, ,
\end{equation}
which is almost a factor of 3 weaker than the 0CV bound. In more realistic conditions we expect a sensible loss of signal from the requirement of displacement, although better background discrimination might be also achieved thanks to the converted photon. We do not even study the 2CV photon category because it is plagued by a very small signal efficiency.  

As a useful input for future more detailed studies, we collect here some considerations about the LHCb reach outside the interval $4.9~{\rm GeV} < m_{\gamma\gamma} < 6.3~{\rm GeV}$: \begin{itemize}
\item[$\diamond$] As far as the signal is concerned, we do not expect a significant drop in the efficiency going at higher invariant masses. As detailed in Appendix~\ref{app:simulations} at higher invariant masses the diphoton final state will be less forward, reducing the geometric acceptance.
However, the decreasing boost of the produced particle is more than compensated by the higher efficiency of the photon $p_T$ cuts.
Practically, the ultimate high mass reach of LHCb is not very relevant for the purposes of discovering new physics, since above 10-20 GeV it is likely to be superseded by the ATLAS/CMS diphoton searches (see \cite{Mariotti:2017vtv} for details).
\item[$\diamond$] 
The most stringent limitation for scanning masses above $\sim$12 GeV at LHCb is the current dynamic range of the ECAL. This range, which depends on the electronics and not on the actual configuration of the detector, limits at the moment reconstructing photons with $E_T$ above $\sim$10 GeV ($\sim$6 GeV at the level of the first level of the software trigger HLT1).
Therefore, a potential increase in the dynamic range of the ECAL after the LHCb Upgrade would be very benificial to increase LHCb's sensitivity to higher masses. For instance, modifying the electronics to increase the range to $15-20$ GeV would be enough to cover all the mass range for which ATLAS and CMS have a poor sensitivity.
\item[$\diamond$] 
As already mentioned, the invariant mass distribution in the 0CV category from the data in Ref.~\cite{Benson:2314368} displays a drop for masses larger than approximately
$m_{\gamma\gamma} \simeq 5.7$ GeV. 
In Appendix~\ref{app:0CVsmooth}, we argue that such drop is a consequence of the use of $2\times2$ ECAL cells to measure the photon energy deposition at HLT1. If our guess is correct there should be another drop of the background at low invariant masses in a region not showed by the plot of Ref.~\cite{Benson:2314368}. 
\item[$\diamond$] Understanding the composition of the diphoton background given in Ref.~\cite{Benson:2314368} would require a detailed MC simulation, including detector effects, which is beyond the scope of this paper. In Appendix~\ref{app:fake_pions} we provide a simple kinematical argument which shows that the background from boosted $\pi^0$ faking photons is likely to dominate over the one from real photons. A categorization of the data in different $\eta$ regions would help suppressing this background at small $\eta$. This could be used to maximize the reach. A quantitative assessment of this is left for future studies. 
\item[$\diamond$] 
The precise assessment of the 1CV limit and sensitivities would require a dedicated search for promptly decaying resonances \emph{without} the requirement of a displaced vertex. In this case one could get an even better reach than the one presented here by combining the 0CV and the 1CV category.

\end{itemize}

We hope that this work could provide enough %theory
motivation to explore further the open issues described above and in general the possibility of performing bump hunts on the top of the diphoton background at such low invariant masses. 

\section{Conclusions}
\label{sec:conclusions}

The LHC has pushed the energy scale of many motivated SM extensions beyond the TeV range. How to experimentally test NP models at and beyond those scales?
A possibility is to look for low energy remnants of such theories, like pseudo-Goldstone bosons (aka ALPs) from an approximate global symmetry.

In Section~\ref{sec:theory} we showed that ALPs with masses and decay constants of interest for flavor factories arise as a solution to the strong CP problem (``heavy QCD axions'') and in frameworks motivated by Dark Matter freeze-out and  the Higgs hierarchy problem (e.g. the SUSY $R$-axion as mediator of DM interactions). These scenarios share the prediction of ALP couplings to gluons and photons, that are currently tested in a particularly poor way for masses below $O(10)$~GeV.

In Section~\ref{sec:bound}, we have used $80$~pb$^{-1}$ of public LHCb data to set a bound on diphoton resonances of $\sigma (pp \to X a(\gamma\gamma)) \lesssim 100$~pb, and we have performed a first study to assess future LHCb sensitivities. This bound is already the strongest existing one on the ALPs discussed above, and 
shows that LHCb has a very promising potential to test unexplored territory of well-motivated BSM extensions. Technical results that might be useful for future LHCb studies are provided in Appendices~\ref{app:fake_pions} and \ref{app:simulations}.
We have also recasted BABAR limits on $\Upsilon \to \gamma a(jj)$ on this model, and estimated the associated future capabilities of Belle-II, finding they would be particularly relevant for masses below $\approx 3$~GeV. These results are summarised in Figure~\ref{fig:ALP}.

Our findings provide a strong motivation to pursue the phenomenological and experimental program of testing this class of ALPs at LHCb and Belle-II, thus enriching the physics case of both machines.

\medskip

\subsection{Acknowledgements}

%\acknowledgements
We thank Sean Benson and Albert Puig Navarro for many useful discussions, in particular about the LHCb note~\cite{Benson:2314368}, and Marco Bonvini for clarifications about ggHiggs. D.R. thanks Simon Knapen for discussion and clarifications on Ref.~\cite{Knapen:2017ebd}. We thank Luca Merlo, Federico Pobbe, Stefano Rigolin and Olcyr Sumensari for pointing out a mistake in the prefactor of eq.~\eqref{eq:Upsilon} in the first version of this paper.
D.R. thanks CERN and the Galileo Galilei Institute for theoretical physics (GGI) for kind hospitality during the completion of this work. F.S. is grateful to the Mainz Institute for Theoretical Physics (MITP), CERN and GGI for kind hospitality at various stages of this work. K.T. thanks MITP for kind hospitality during the completion of this work.
\medskip

{\footnotesize
\noindent Funding and research infrastructure acknowledgements: 
\begin{itemize}
\item[$\ast$] X.C.V. is supported by MINECO through the Ram\'on y Cajal program RYC-2016-20073;
\item[$\ast$] A.M. is supported by the SRP High Energy Physics and the Research Council of the Vrije Universiteit Brussel;
A.M is also supported by FWO under the EOS-BE.H project n. 30820817
\item[$\ast$] D.R. is supported in part by the National Science Foundation under Grant No. NSF PHY17-48958;
\item[$\ast$] F.S is supported in part by a {\sc Pier} Seed Project funding (Project ID PIF-2017-72);
\item[$\ast$] K.T. is supported  by his start-up funding at Florida State University.
\end{itemize}
}

\appendix

\section{More on the Signal}
\label{app:signal}

\begin{figure*}[th]
\includegraphics[width=0.49\textwidth]{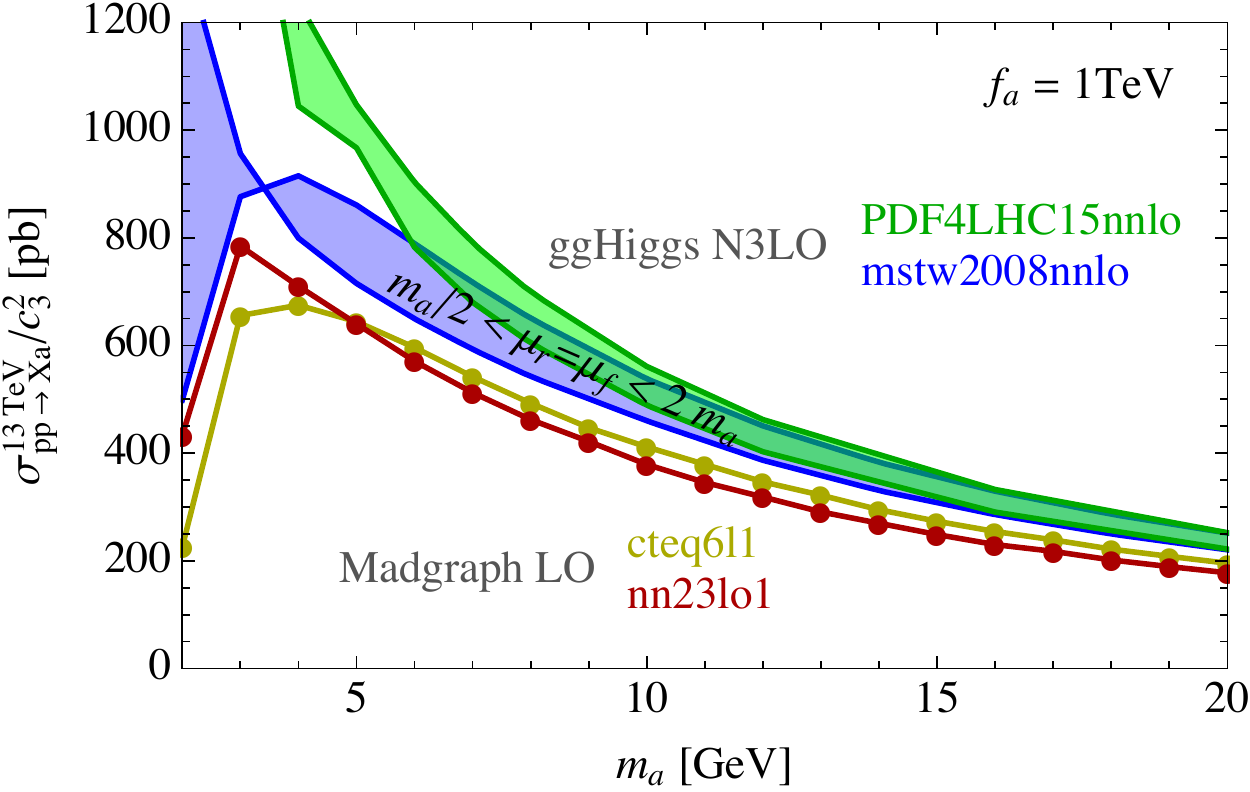} \hfill
\includegraphics[width=0.49\textwidth]{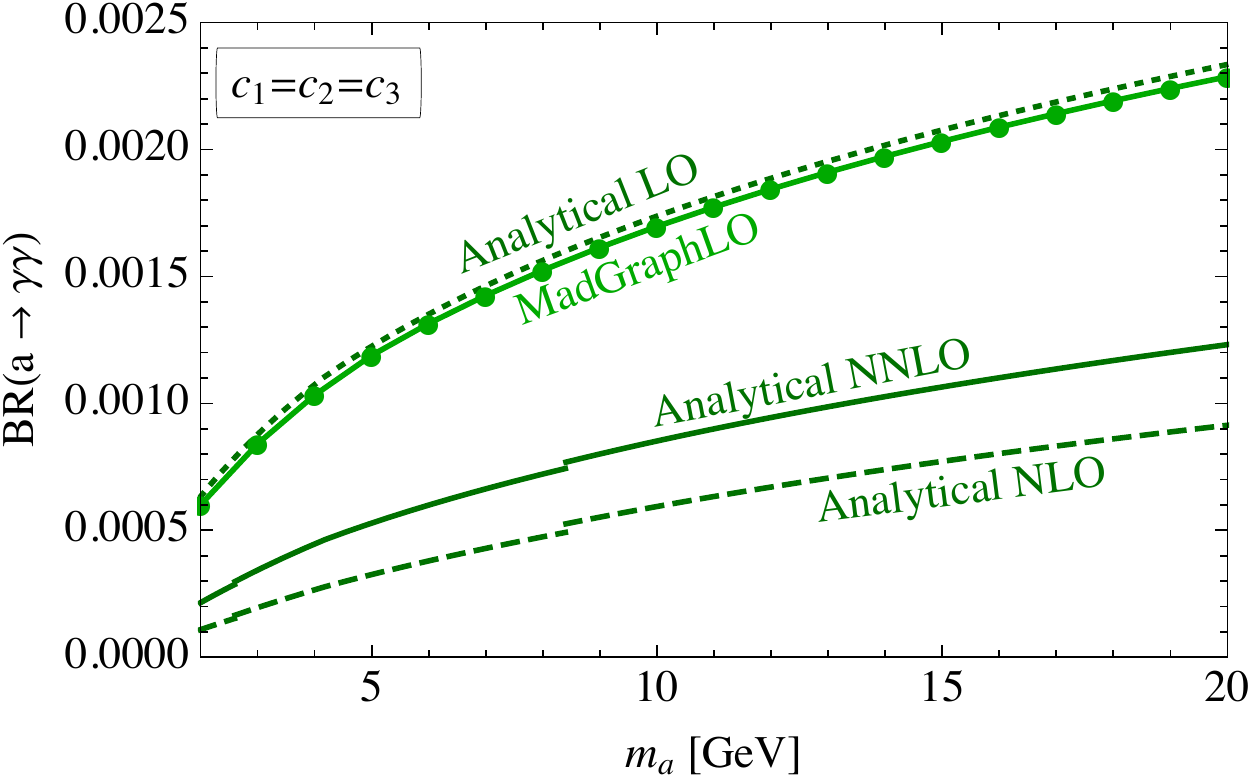}
\caption{\label{fig:Signal} Left: Production cross section of an ALP coupled to $G\tilde G$, as determined with MadGraphLO and with ggHiggs at N3LO, for various choices of the pdf sets, fixing $f = 1$~TeV and $c_3=1$. For ggHiggs we display the band enclosed by $\mu_f = \mu_r = m_a/2$ and $\mu_f = \mu_r = 2\,m_a$.
Right: ALP branching ratio into diphoton at LO, NLO and NNLO, and from MadGraphLO.}
\end{figure*}

We compute the gluon fusion production cross section at N3LO using ggHiggs v4~\cite{Ball:2013bra,Bonvini:2014jma,Bonvini:2016frm,Ahmed:2016otz} and at LO using MadGraphLO. We compare the two predictions in Figure~\ref{fig:Signal} left, for different choices of the pdf sets, and rescaling the ggHiggs cross section using that $c_3/f = 1/(\sqrt{2}\,v)$ with $v \simeq 246$~GeV (anomaly coefficient coming from a top loop).
%We find that matching or not to up to two extra jets have an impact at the percent level on the determination of the signal from Madgraph plus Pythia.
The agreement between these determinations goes from the 20\% level at $m_a = 20$~GeV, down to a factor of 2 and worse for $m_a \leq 4$~GeV. We mention that at such low values the ggHiggs output should be taken with extra care, as it also yields some negative LO and NLO cross sections.
This comparison underlines the need for a more precise determination of the production cross section, especially for ALP masses below 5 GeV or so. This task goes however beyond the purpose of this paper.
We use the ggHiggs prediction with the {\tt mstw2008nnlo} pdf set for all the LHC phenomenology in Section~\ref{sec:ALPs}. 

Coming now to the ALP branching ratios, we use the NNLO QCD correction to the width of a pseudoscalar into gluons from~\cite{Chetyrkin:1998mw}. In the notation of Eq.~(\ref{eq:widths}), it reads $K_{gg} = 1+\frac{\alpha_s^{(5)}}{\pi}\,E_A\, + \big(\frac{\alpha_s^{(5)}}{\pi}\big)^2\,E_A\,\big(\frac{3}{4}\,E_A  + \frac{\beta_1}{\beta_0}\big)$, where $E_A = \frac{97}{4} - \frac{7}{6} N_f$, $\beta_0 = \frac{11}{4} - \frac{N_f}{6}$, $\beta_1= \frac{51}{8} - \frac{19}{24} N_f$.
In Figure~\ref{fig:Signal} right we plot the resulting diphoton branching ratio together with its NLO and LO value and with the one given by Madgraph. NNLO corrections to the diphoton branching ratio reduce its LO value by a factor of $\simeq 2$, over the whole mass range we consider.
We use the NNLO expression for all the limits and sensitivities described in Section~\ref{sec:ALPs}.

%\FS{ggHiggs generates the coupling to gluons from the one to top, but it does not report anywhere the normalisation it uses for the coupling to the top of the pseudoscalar.I think it uses $y_t\,a\,\bar{t} \gamma_5 t$, because this gives a $\Gamma(a \to gg) = \frac{9}{4}\Gamma(h\to gg)$, where $h$ is the SM Higgs (see e.g. the Djouadi MSSM review, from page 90 on).Anyway, in producing the cross sections from ggHiggs I used that the anomaly generated from the coupling $y_t\,a\,\bar{t} \gamma_5 t$ is \beq  \frac{c_{3,t}}{f} = \frac{1}{2}\frac{1}{v},\qquad v=176~{\rm GeV} \label{eq:anomaly_yt} \eeq
%and the offset could well be due to a factor of 2 in the above equation.
%COULD ANYBODY CHECK Eq.~(\ref{eq:anomaly_yt})?}

\section{$m_{\gamma\gamma}$ distribution of the 0CV category}
\label{app:0CVsmooth}
\begin{figure*}[th]
\includegraphics[width=0.49\textwidth]{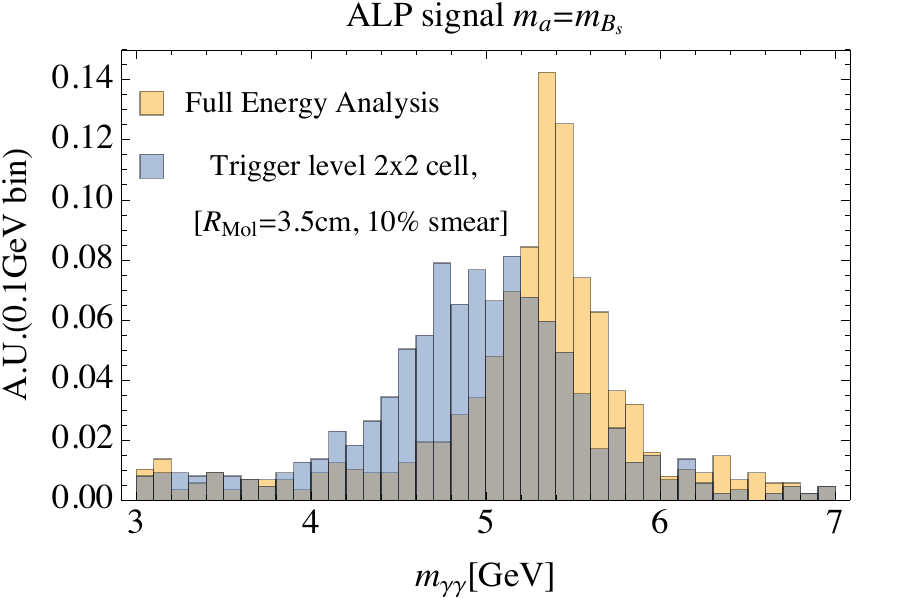} \hfill
\includegraphics[width=0.49\textwidth]{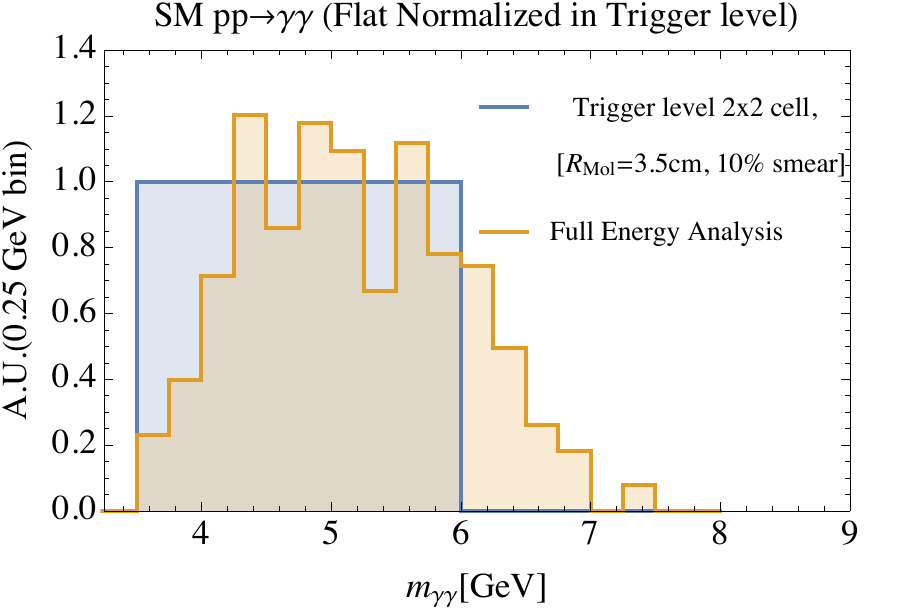}
\caption{\label{fig:2x2study} 
Left: ALP signal event with $m_a=m_{B_s}$ in diphoton invariant mass $m_{\gamma\gamma}^{\rm full}$ (yellow) and trigger level diphoton invariant mass $m_{\gamma\gamma}^{\rm trigger}$ (blue). Fraction of energy, $E^{(2\times2)}_{\gamma}=E^{\rm full}_{\gamma}\min[1,{\cal P}_{\rm normal}(\mu=0.95,\sigma=0.1)]$, is used for the calculation of trigger level invariant mass. 
Right: SM diphoton event with a cut, $3.5~{\rm GeV}<m_{\gamma\gamma}^{\rm trigger }<6~{\rm GeV}$, in $m_{\gamma\gamma}^{\rm full}$ (yellow) and in $m_{\gamma\gamma}^{\rm trigger}$ (blue). To demonstrate bin migration effect, the distribution in $m_{\gamma\gamma}^{\rm trigger}$ is flat normalized. 
}
\end{figure*}

In our analysis, we assume the background yield to be roughly constant with respect to the diphoton invariant mass even outside the mass range reported in Ref.~\cite{Benson:2314368}.
This is to provide an order-of-magnitude estimate of the background for the LHCb sensitivities to ALPs.
The flatness of the data is actually seen in the 1CV and 2CV categories of Fig.~4(b--c) of Ref.~\cite{Benson:2314368}. However, in the 0CV category (Fig.~4(a)), a kink is observed at large invariant masses. In what follows, we argue that this is an artifact due to the trigger level invariant mass cut.   

%We examined two possibilities: (1) less ECAL cells used for photon energy lowers the invariant mass; (2) the approximation for diphoton angle affects the distribution, and concluded (1) is the reason for the kink while (2) has negligible impact.
In the invariant mass calculation at the trigger level of the 0CV category, two approximations are employed to speed up the calculation: 1) the photon energy is calculated from the energy deposition in $2\times2$ ECAL cells, 2) the mass formula takes into account only the leading order of the diphoton opening angle, $m_{\gamma\gamma}^{\rm trigger }=\sqrt{E^{(2\times2)}_{\gamma_1}E^{(2\times2)}_{\gamma_2}}\Delta\theta_{\gamma_1\gamma_2}$. We examined these two approximation and concluded that 1) could be the reason for the kink.

It is easy to show that the approximate mass formula is equivalent to the full mass formula with  ${\cal O}(0.01)$ accuracy. This comes from the fact that the diphoton events within the LHCb fiducial volume  have a small opening angle $\Delta\theta_{\gamma_1\gamma_2}={\cal O}(0.1)$, after the $E_T$ trigger cuts are imposed.  On the other hand one  needs to use $3\times3$ cells to capture full energy deposit of a photon, so the information based on $2\times 2$ cells  underestimates the photon energy, which leads to the lower invariant mass, $m_{\gamma\gamma}^{\rm trigger }<m_{\gamma\gamma}^{\rm full}$.
Because the first invariant mass cut is made at the trigger level, $3.5~{\rm GeV}<m_{\gamma\gamma}^{\rm trigger }<6~{\rm GeV}$, bins with a given $m_{\gamma\gamma}^{\rm trigger }$ migrate to bins with $m_{\gamma\gamma}^{\rm full} > m_{\gamma\gamma}^{\rm trigger}$.This could explain why the reduction of the yield appears above $m_{\gamma\gamma}^{\rm full}\sim 6$~GeV.
This argument is confirmed by Fig.~1 bottom of~\cite{Benson:2314368}, that shows how the trigger level mass distribution of the $B_s$ signal shifts to higher values of off-line invariant mass. 

We further validate the argument modelling the energy smearing of the LHCb ECAL. For simplicity, we focus on the inner ECAL and approximate 2$\times$2 cells as a circle of radius 4~cm. Because the Moli\`ere radius of a photon in the LHCb ECAL is 3.5cm,\footnote{Inside
the Moli\`ere radius, the energy deposit into the corresponding area is 90\% of the total energy on average.}
the energy deposit inside the 2$\times$2 cells is expected to be 95\% of the total energy deposit on average. In order to model a realistic environment we include a stochastic gaussian smearing from the average value. We choose a standard deviation of 10\%\footnote{$E^{(2\times2)}_{\gamma}=E^{\rm full}_{\gamma}\min[1,{\cal P}_{\rm normal}(\mu=0.95,\sigma=0.1)]$} such that the shift of the signal at $m_a=m_{B_s}$ reproduces Fig.~1 bottom of~\cite{Benson:2314368}. This result is shown in Fig.~\ref{fig:2x2study} left.
Then, we use the same prescription for the background-like events. The result is shown in Fig.~\ref{fig:2x2study} right. The invariant mass distribution in terms of $m_{\gamma\gamma}^{\rm trigger }$ is normalized to be rectangular after the invariant mass cut. When the same dataset is plotted in terms of $m_{\gamma\gamma}^{\rm full}$ we can see that a kink is induced.

\section{Details on the LHCb Calorimeter}
\label{app:fake_pions}

The Electromagnetic Calorimeter (ECAL) of LHCb has three layers with different granularities and is placed vertically with respect to the beam axis at $z_{\rm Ecal}$=12.52~m away from the collision point. The ECAL square cells have side lengths of   $\Delta x_{\rm cell}$=4.04, 6.06cm, 12.12cm for inner, middle, and outer layer, respectively \cite{Aaij:2014jba}.
The photon reconstruction algorithm uses patterns of $3\times 3$ cells in each layer. Therefore, the inner layer, where most of the energy is expected to be deposited, has the best angular resolution. 

\paragraph{Invariant mass resolution}
%We can determine the invariant mass resolution.
The invariant mass can be written as 
\begin{equation}
m_{\gamma\gamma}^2=2E_{\gamma_1} E_{\gamma_2}(1-\cos\theta_{\gamma\gamma})\ ,
\end{equation}
where $E_{\gamma_{1,2}}$ are the energies of the two photons and $\theta_{\gamma\gamma}$ is the angular separation between them. Using the above formula, we can relate the invariant mass smearing to the photon energy smearing and the ECAL granularity

\begin{align}
&\frac{\delta m_{\gamma\gamma}}{m_{\gamma\gamma}}
\simeq \frac{1}{2}
\frac{\delta m^2_{\gamma\gamma}}{m^2_{\gamma\gamma}}
=
\frac{1}{2}\left(\frac{\delta E_{\gamma_1}}{E_{\gamma_1}}\oplus \frac{\delta E_{\gamma_2}}{E_{\gamma_2}}
\oplus \frac{\sin\theta_{\gamma\gamma}\delta\theta}{1-\cos\theta_{\gamma\gamma}}\right)
\nonumber\\
&\simeq  
\frac{1}{\sqrt{2}}\frac{\delta E_{\gamma}}{E_{\gamma}}
\oplus \frac{\delta\theta}{\theta_{\gamma\gamma}}
=6.4\%\sqrt{\frac{{\rm GeV}}{E_\gamma}}\oplus 0.6\%\oplus 0.3\%\frac{E_\gamma}{m_{\gamma\gamma}}\ .
\end{align}
In the second line we assumed for simplicity $E_{\gamma_1}\simeq E_{\gamma_2}\simeq E_{\gamma}$ and approximated our result at the first order in $\theta\ll1$.
To obtain the second expression in the second line, we used the LHCb ECAL energy resolution $\delta E/E \simeq 9\% \sqrt{{\rm GeV}/E} \oplus 0.8\%$ reported in Ref.~\cite{Perret:2014owa} and the granularity of the inner layer of the ECAL $\delta\theta=\Delta x_{\rm cell}/z_{\rm Ecal}\simeq 0.003$.
Moreover, we have approximated $\theta_{\gamma\gamma}\simeq m_{\gamma\gamma}/E_\gamma$ to get an expression of the typical energy smearing as a function of the typical photon energy.  In computing the invariant mass resolution in the text, we take $E_{\gamma}=50\rm GeV$. We believe this is a realistic benchmark value for this analysis because $E_\gamma = E_{T\gamma}\cosh\eta$ and the LHCb analysis in Ref.~\cite{Benson:2314368} imposes $E_{T\gamma} > 3.5$~GeV and $E_{T\gamma 1}+E_{T\gamma 2} > 8$~GeV on $2\times2$ cell clusters.

\paragraph{Background from $\pi^0$ faking single photon}
One of the advantages to study low mass diphoton resonances at LHCb is that low energy fake photons from QCD   can be  distinguished from real photon candidates. 
Here we focus on fake photons from $\pi^0$ decays whose collimated diphoton decay can mimick a single photon candidate. 

\begin{figure}[t]
\includegraphics[width=0.47\textwidth]{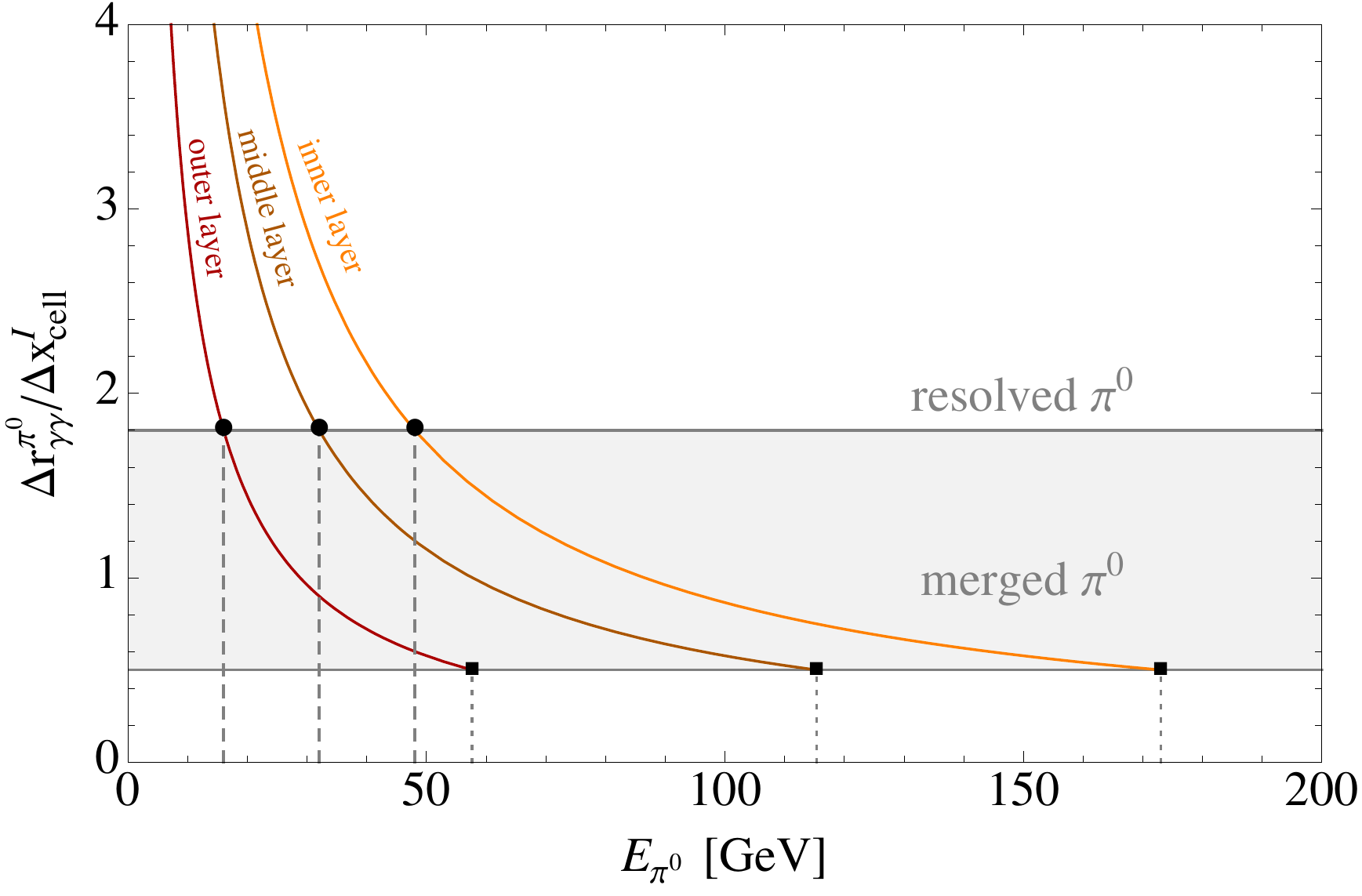}
%\hfill \includegraphics[width=0.47\textwidth]{Figures/Ephoton_vs_mALP.pdf}
\caption{\label{fig:piBG} 
%Left:
Separation of photon pairs from $\pi^0$ decay as a function of the pion total energy $E_{\pi^0}$.
If this photon pair is misidentified as a single (fake) photon, $E_{\pi^0}$ is the energy of the fake photon. Cases of inner, middle, and outer layers are plotted in blue, red and magenta respectively. 
%Right: Total energy of each of the signal photons, as a function of the mass of the resonance, and for different values of the pseudorapidity and transverse momentum of the resonance (all satisfying $p_{Ta} > 2\, E_{T\gamma}^\text{cut} = 7$~GeV, and approximating $\eta_{\gamma1} = \eta_{\gamma2}$ for simplicity). The energy for which the inner ECAL would safely reject a pion faking a photon is shown as a dashed green line for reference.
}
\end{figure}

Photon pairs from $\pi^0$ decay have angular separation
$\theta^{\pi^0}_{\gamma\gamma}\simeq {m_{\pi^0}}/{E_{\gamma}}\simeq {2m_{\pi^0}}/{E_{\pi^0}}$. The corresponding separation on a given ECAL layer is then
\begin{align}
\Delta r^{\pi^0}_{\gamma\gamma} \simeq z_{\rm Ecal}\theta^{\pi^0}_{\gamma\gamma}\simeq \frac{2 z_{\rm Ecal }m_{\pi^0} }{E_{\pi^0}}\ .
\end{align}
If the $\pi^0$ is very energetic, the diphoton separation $\Delta r^{\pi^0}_{\gamma\gamma}$ is smaller than a single cell size and the object is mostly misidentified as a single photon candidate of energy $E_{\pi^0}$. Viceversa, when a pion is less energetic and the diphoton separation is large, $\Delta r^{\pi^0}_{\gamma\gamma}> {\cal O}(2)\Delta x_{cell}$, two photon clusters are separately formed and a pion is {\it resolved}. In a regime where $1.8\Delta x_{cell}>\Delta r^{\pi^0}_{\gamma\gamma}\gtrsim 0.5 x_{cell}$, the shower shape information makes a single energy cluster identified as a $\pi^0$, which is called {\it merged $\pi^0$} \cite{Deschamps:2003lja}. The identification efficiency using both {\it resolved} and {\it merged $\pi^0$} is $\cal{O}$(50\%) for $p_{T{\pi^0}}\lesssim 10$~GeV (Fig.~21 left of Ref.~\cite{Deschamps:2003lja}). 
As shown in Fig.~\ref{fig:piBG}, the final energy thresholds vary depending on the ECAL layer. For example, in the inner ECAL
diphotons with $E_{\pi^0}<28$~GeV corresponding to a large separation of $\Delta r^{\pi^0}_{\gamma\gamma}>3\Delta x_{cell}$ can be reconstructed as resolved $\pi^0$s, while the ones with $46~{\rm GeV}<E_{\pi^0}\lesssim 160$~GeV could be seen as merged $\pi^0$s. 

The planned LHCb $B_s\to \gamma\gamma$ analysis uses a photon energy threshold of $E_{T\gamma}>$3.5~GeV which corresponds to $E_\gamma=13\, (260)$~GeV at $\eta=$2 (5). Comparing with the threshold determined above for the pions to be detected as fake photons, one learns that i) the background to the current search contains a non-negligible amount of fake photons; ii) a categorization in $\eta$  of the data could help in reducing photon fakes. 

%\DR{As a further check we simulated...}

%We show the dependence of the photon energy on the resonance mass in the right-hand panel of Figure~\ref{fig:piBG}. This appears to be a much milder dependence than the one on the analysis cuts, at least for $m_a$ within a few tens of GeV, and therefore into a mass regime where ATLAS and CMS diphoton searches are anyway expected to perform better than LHCb ones. Note that in the right-hand panel of Figure~\ref{fig:piBG} we have approximated the two resonant photons to have the same $\eta$ for simplicity.

%[what follows is the previous version, which seems not correct to me. The ALP mass does not play an important role, and indeed I would also show the other figure]} If this threshold is kept, one problem to extend this analysis to higher mass resonance with $m_a\gg m_{B_s}$ is that at least one photon (or even two) would be in a regime of high rapidity or high transverse momentum where photon fakes are irreducible.  This observation could cause a degradation of the LHCb sensitivity at higher invariant masses, because of the limited rapidity coverage of LHCb ($2<\eta<5$) and the condition $m_a\simeq \Delta\eta_{\gamma\gamma}\sqrt{p_{T{\gamma_1}}p_{T{\gamma_2}}}$ \FS{Check this formula and put it above}. A $\eta$ categorization could help improving this issue. \DR{do you agree?} 

\section{Signal Acceptance and Efficiency}
\label{app:simulations}

In this Appendix we discuss the strategy that we adopted to estimate the acceptance and efficiency of the signal. As mentioned in the main text, 
we eventually consider a constant value for the product of acceptance times efficiency on the mass range of interest for this paper.
As reference value, we have chosen the one at the invariant mass of $5$ GeV, corresponding to the $B_s$ signal considered in the LHCb note 
\cite{Benson:2314368}.

In order to estimate the acceptance and efficiency of the signal at LHCb, we implement the axion model in 
FeynRules~\cite{Alloul:2013bka}, we generate events with MadgraphLO\_v2\_6~\cite{Alwall:2011uj,Alwall:2014hca} and shower them with Pythia~8.1~\cite{Sjostrand:2006za,Sjostrand:2007gs}, matching up to 1 extra jets~\cite{Alwall:2007fs}.
We then perform a simple analysis of the resulting samples using MadAnalysis5 \cite{Conte:2012fm}. Note that the signal events which are inside the acceptance of LHCb contain topologies where the axion has acquired a significant longitudinal boost,
 without the need of extra hard radiation.
 As a consequence the signal efficiency is essentially not changed by including extra jets (the minimal $E_T$ cuts of
 the LHCb selection can be satisfied with just a small transverse boost).
 This has to be contrasted with the low invariant mass searches at ATLAS/CMS where the recoil of the resonance against the extra jet 
 increases the signal efficiency of the $p_T$ cuts significantly, as it was shown in Ref.~\cite{Mariotti:2017vtv}.

In Table \ref{tab:efficie} we report the acceptance and the efficiency that we find in the mass range $5-15$ GeV by following the selection cuts of refs. \cite{Benson:2314368}, that is 
\begin{eqnarray}
&&\mathcal{A}: \quad 2 <\eta(\gamma) <5 \label{eq:acce} \\
&& \epsilon: 
\left \{
\begin{array}{l}
E_T(\gamma) > 3.5 \text{ GeV},  
\vspace{.15cm}
\\
E_T^{\gamma_1}+E_T^{\gamma_2} > 8 \text{ GeV}
\vspace{.15cm}
\\
p_T({\gamma_1 \gamma_2}) > 2 \text{ GeV} 
\vspace{.15cm}
\\
\end{array}
\right . 
\label{eq:cuts}
\end{eqnarray}
We first observe that the value we find for the product $\mathcal{A} \times \epsilon$, though in the same ballpark than the number reported by the
LHCb note (see eqn.\eqref{eq:bench}), differs by around a factor of $2$ on the case of $m_a = 5$ GeV.
In order to check wheter the discrepancy could be caused by detector effects, we also processed the same samples using Delphes as fast LHCb detector simulator, but we did
not find a substantial improvement in the agreement. 

However, besides the discrepancy on the benchmark of $5$ GeV, 
our simple analysis provides indications on what could be the expected product of $\mathcal{A} \times \epsilon$ for the selection cuts \eqref{eq:cuts} for different mass values.
As one can observe from Table \ref{tab:efficie}, increasing the mass of the axion the acceptance generically decreases. This is due to the fact that a heavier resonance will more likely be produced with less boost on the longitudinal axis, and hence the resulting photons will be less into the forward region which is covered by the LHCb detector.
On the other hand, for larger values of the axion mass the outgoing photons will be more energetic and will more likely pass the energy and $p_T$ cuts, hence
resulting in an increase in the signal efficiency. The combination of these two effects result in a product of acceptance times efficiency which actually slightly grows along the mass interval $5 -15$ GeV, but does not changes significantly.
This justifies the simplified choice that we have adopted in the main part of the paper.

\begin{table}[h]
\begin{tabular}{|c||c|c|c|c|c|c|}
\hline
$m_a$[GeV] & 5 & 7 & 9 & 11 & 13 & 15 \\
\hline
$\mathcal{A}$ &0.15 &0.15 & 0.14 &0.13 & 0.12 &0.12\\
\hline
$\epsilon$ &0.26 &0.36& 0.46&0.64&0.72&0.81\\
\hline
\end{tabular}
\caption{Acceptance \eqref{eq:acce} and efficiency \eqref{eq:cuts} for the axion signal in the LHCb anaysis, for different mass values.
\label{tab:efficie}}
\end{table}

\bibliography{Diphoton_LHCb}
\end{document}